



\font\bigbold=cmbx12
\font\ninerm=cmr9      \font\eightrm=cmr8    \font\sixrm=cmr6
\font\fiverm=cmr5
\font\ninebf=cmbx9     \font\eightbf=cmbx8   \font\sixbf=cmbx6
\font\fivebf=cmbx5
\font\ninei=cmmi9      \skewchar\ninei='177  \font\eighti=cmmi8
\skewchar\eighti='177  \font\sixi=cmmi6      \skewchar\sixi='177
\font\fivei=cmmi5
\font\ninesy=cmsy9     \skewchar\ninesy='60  \font\eightsy=cmsy8
\skewchar\eightsy='60  \font\sixsy=cmsy6     \skewchar\sixsy='60
\font\fivesy=cmsy5     \font\nineit=cmti9    \font\eightit=cmti8
\font\ninesl=cmsl9     \font\eightsl=cmsl8
\font\ninett=cmtt9     \font\eighttt=cmtt8
\font\tenfrak=eufm10   \font\ninefrak=eufm9  \font\eightfrak=eufm8
\font\sevenfrak=eufm7  \font\fivefrak=eufm5
\font\tenbb=msbm10     \font\ninebb=msbm9    \font\eightbb=msbm8
\font\sevenbb=msbm7    \font\fivebb=msbm5
\font\tenssf=cmss10    \font\ninessf=cmss9   \font\eightssf=cmss8
\font\tensmc=cmcsc10

\font\tensvec=cmmib10
\font\sevensvec=cmmib10 at 7pt
\font\fivesvec=cmmib10 at 5 pt
\def\svec#1{{\mathchoice%
  {\hbox{$\displaystyle\textfont1=\tensvec%
                       \scriptfont1=\sevensvec%
                       \scriptscriptfont1=\fivesvec#1$}}%
  {\hbox{$\textstyle\textfont1=\tensvec%
                    \scriptfont1=\sevensvec%
                    \scriptscriptfont1=\fivesvec#1$}}%
  {\hbox{$\scriptstyle\scriptfont1=\sevensvec%
                      \scriptscriptfont1=\fivesvec#1$}}%
  {\hbox{$\scriptscriptstyle\scriptscriptfont1=\fivesvec#1$}}%
}}

\newfam\bbfam   \textfont\bbfam=\tenbb \scriptfont\bbfam=\sevenbb
\scriptscriptfont\bbfam=\fivebb  \def\Bbb{\fam\bbfam}
\newfam\frakfam  \textfont\frakfam=\tenfrak \scriptfont\frakfam=%
\sevenfrak \scriptscriptfont\frakfam=\fivefrak  \def\frak{\fam\frakfam}
\newfam\ssffam  \textfont\ssffam=\tenssf \scriptfont\ssffam=\ninessf
\scriptscriptfont\ssffam=\eightssf  
\def\smc{\tensmc}

\def\eightpoint{\textfont0=\eightrm \scriptfont0=\sixrm
\scriptscriptfont0=\fiverm  \def\rm{\fam0\eightrm}%
\textfont1=\eighti \scriptfont1=\sixi \scriptscriptfont1=\fivei
\def\oldstyle{\fam1\eighti}\textfont2=\eightsy
\scriptfont2=\sixsy \scriptscriptfont2=\fivesy
\textfont\itfam=\eightit         \def\it{\fam\itfam\eightit}%
\textfont\slfam=\eightsl         \def\sl{\fam\slfam\eightsl}%
\textfont\ttfam=\eighttt         \def\tt{\fam\ttfam\eighttt}%
\textfont\frakfam=\eightfrak     \def\frak{\fam\frakfam\eightfrak}%
\textfont\bbfam=\eightbb         \def\Bbb{\fam\bbfam\eightbb}%
\textfont\bffam=\eightbf         \scriptfont\bffam=\sixbf
\scriptscriptfont\bffam=\fivebf  \def\bf{\fam\bffam\eightbf}%
\abovedisplayskip=9pt plus 2pt minus 6pt   \belowdisplayskip=%
\abovedisplayskip  \abovedisplayshortskip=0pt plus 2pt
\belowdisplayshortskip=5pt plus2pt minus 3pt  \smallskipamount=%
2pt plus 1pt minus 1pt  \medskipamount=4pt plus 2pt minus 2pt
\bigskipamount=9pt plus4pt minus 4pt  \setbox\strutbox=%
\hbox{\vrule height 7pt depth 2pt width 0pt}%
\normalbaselineskip=9pt \normalbaselines \rm}

\def\ninepoint{\textfont0=\ninerm \scriptfont0=\sixrm
\scriptscriptfont0=\fiverm  \def\rm{\fam0\ninerm}\textfont1=\ninei
\scriptfont1=\sixi \scriptscriptfont1=\fivei \def\oldstyle%
{\fam1\ninei}\textfont2=\ninesy \scriptfont2=\sixsy
\scriptscriptfont2=\fivesy
\textfont\itfam=\nineit          \def\it{\fam\itfam\nineit}%
\textfont\slfam=\ninesl          \def\sl{\fam\slfam\ninesl}%
\textfont\ttfam=\ninett          \def\tt{\fam\ttfam\ninett}%
\textfont\frakfam=\ninefrak      \def\frak{\fam\frakfam\ninefrak}%
\textfont\bbfam=\ninebb          \def\Bbb{\fam\bbfam\ninebb}%
\textfont\bffam=\ninebf          \scriptfont\bffam=\sixbf
\scriptscriptfont\bffam=\fivebf  \def\bf{\fam\bffam\ninebf}%
\abovedisplayskip=10pt plus 2pt minus 6pt \belowdisplayskip=%
\abovedisplayskip  \abovedisplayshortskip=0pt plus 2pt
\belowdisplayshortskip=5pt plus2pt minus 3pt  \smallskipamount=%
2pt plus 1pt minus 1pt  \medskipamount=4pt plus 2pt minus 2pt
\bigskipamount=10pt plus4pt minus 4pt  \setbox\strutbox=%
\hbox{\vrule height 7pt depth 2pt width 0pt}%
\normalbaselineskip=10pt \normalbaselines \rm}

\global\newcount\secno \global\secno=0 \global\newcount\meqno
\global\meqno=1 \global\newcount\appno \global\appno=0
\newwrite\eqmac \def\romappno{\ifcase\appno\or A\or B\or C\or D\or
E\or F\or G\or H\or I\or J\or K\or L\or M\or N\or O\or P\or Q\or
R\or S\or T\or U\or V\or W\or X\or Y\or Z\fi}
\def\eqn#1{ \ifnum\secno>0 \eqno(\the\secno.\the\meqno)
\xdef#1{\the\secno.\the\meqno} \else\ifnum\appno>0
\eqno({\rm\romappno}.\the\meqno)\xdef#1{{\rm\romappno}.\the\meqno}
\else \eqno(\the\meqno)\xdef#1{\the\meqno} \fi \fi
\global\advance\meqno by1 }

\global\newcount\refno \global\refno=1 \newwrite\reffile
\newwrite\refmac \newlinechar=`\^^J \def\ref#1#2%
{\the\refno\nref#1{#2}} \def\nref#1#2{\xdef#1{\the\refno}
\ifnum\refno=1\immediate\openout\reffile=refs.tmp\fi
\immediate\write\reffile{\noexpand\item{[\noexpand#1]\ }#2\noexpand%
\nobreak.} \immediate\write\refmac{\def\noexpand#1{\the\refno}}
\global\advance\refno by1} \def\semi{;\hfil\noexpand\break ^^J}
\def\nl{\hfil\noexpand\break ^^J} \def\refn#1#2{\nref#1{#2}}

\def\ijmp#1#2#3{{\it Int.\ J.\ Mod.\ Phys.}\ {\bf A{#1}} ({#2}) #3}

\def\mplA#1#2#3{{\it Mod.\ Phys.\ Lett.}\ {\bf A{#1}} ({#2}) #3}

\def\pl#1#2#3{{\it Phys.\ Lett.}\ {\bf B{#1}} ({#2}) #3}

\def\prD#1#2#3{{\it Phys.\ Rev.}\ {\bf D{#1}} ({#2}) #3}

\newif\iftitlepage \titlepagetrue \newtoks\titlepagefoot
\titlepagefoot={\hfil} \newtoks\otherpagesfoot \otherpagesfoot=%
{\hfil\tenrm\folio\hfil} \footline={\iftitlepage\the\titlepagefoot%
\global\titlepagefalse \else\the\otherpagesfoot\fi}

\def\abstract#1{{\parindent=30pt\narrower\noindent\ninepoint\openup
2pt #1\par}}

\newcount\notenumber\notenumber=1 \def\note#1
{\unskip\footnote{$^{\the\notenumber}$} {\eightpoint\openup 1pt #1}
\global\advance\notenumber by 1}

\def\today{\ifcase\month\or January\or February\or March\or
April\or May\or June\or July\or August\or September\or October\or
November\or December\fi \space\number\day, \number\year}

\def\pagewidth#1{\hsize= #1}  \def\pageheight#1{\vsize= #1}
\def\hcorrection#1{\advance\hoffset by #1}
\def\vcorrection#1{\advance\voffset by #1}

\pageheight{23cm}
\pagewidth{15.7cm}
\hcorrection{-1mm}
\magnification= \magstep1
\parskip=5pt plus 1pt minus 1pt
\tolerance 8000
\def\bsk{\baselineskip= 14.5pt plus 1pt minus 1pt}
\bsk

\font\extra=cmss10 scaled \magstep0  \setbox1 = \hbox{{{\extra R}}}
\setbox2 = \hbox{{{\extra I}}}       \setbox3 = \hbox{{{\extra C}}}
\setbox4 = \hbox{{{\extra Z}}}       \setbox5 = \hbox{{{\extra N}}}





\def\frac#1#2{{#1\over#2}}

\def\pmb#1{\setbox0=\hbox{$#1$} \kern-.025em\copy0\kern-\wd0
    \kern.05em\copy0\kern-\wd0 \kern-.025em\raise.0433em\box0 }

\def\ve{\vfill\eject}


\def\bb{{\svec b}}
\def\bE{{\svec E}}
\def\bB{{\svec B}}
\def\bD{{\svec D}}
\def\bH{{\svec H}}
\def\bk{{\svec k}}
\def\br{{\svec r}}
\def\bd{{\svec d}}
\def\bh{{\svec h}}
\def\bb{{\svec b}}
\def\be{{\svec e}}

\def\btheta{{\svec \theta}}
\def\bkp{{\svec \kappa}}
\def\bve{{\svec \varepsilon}}


{

\refn\Snyder
{H. Snyder, 
{\it Phys. Rev.} {\bf 71} (1947) 38}

\refn\SW
{N. Seiberg and E. Witten,
{\it JHEP} {\bf 9909} (1999) 032, hep-th/9908142}

\refn\GJPP
{Z. Guralnik, R. Jackiw, S.Y.Pi and A.P. Polychronakos,
\pl{517}{2001}{450}}

\refn\Cai
{R.G. Cai, \pl{517}{2001}{457}, hep-th/0106047}

\refn\BCCMV
{G. Berrino, S.L. Cacciatori, A. Celi, L. Martucci, A. Vicini,
\prD{67}{2003}{065021}, hep-th/0210171}

\refn\Banerjee
{R. Banerjee, 
\prD{67}{2003}{105002}, hep-th/0210259}

\refn\GRS
{O. Ganor, G. Rajesh and S. Sethi,
\prD{62}{2000}{125008}, hep-th/0005046}

\refn\RU
{S.-J. Rey and R. von Unge,
\pl{499}{2001}{215}, hep-th/0007089}

\refn\DY
{O. Dayi and B. Yapiskan,
{\it JHEP} {\bf 10} (2002) 022, hep-th/0208043}

\refn\BGPSW
{A. Bichl, J. Grimstrup, L. Popp, M. Schweda and R. Wulkenhaar,
\ijmp{17}{2002}{2219}, hep-th/0102044}

\refn\Aschieri
{P. Aschieri,
\mplA{16}{2001}{163}, hep-th/0103150}

\refn\GZ
{M. Gaillard and B. Zumino,
Nucl. Phys. {\bf B193} (1981) 221}

\refn\Lozano
{Y. Lozano, 
\pl{399}{1997}{233}, hep-th/9701186}

}



\pageheight{23cm}
\pagewidth{15.7cm}
\hcorrection{0mm}
\magnification= \magstep1
\def\bsk{%
\baselineskip= 16.8pt plus 1pt minus 1pt}
\parskip=5pt plus 1pt minus 1pt
\tolerance 6000




\hfill 
{KEK-TH-898}
\vskip -4pt 
\hfill 
\phantom{quant-ph/0207xxx}

\vskip 50pt

{\baselineskip=18pt

\centerline{\bigbold
Duality Symmetry and Plane Waves}
\centerline{\bigbold
in Non-Commutative Electrodynamics}

\vskip 30pt

\centerline{\smc
Yasumi Abe\footnote{${}^{*}$}
{\eightpoint email:\ yasumi@post.kek.jp},
\quad
Rabin Banerjee\footnote{${}^\dagger$}
{\eightpoint On leave from S.N. Bose Natl. Ctr. for Basic Sciences,
Calcutta, India; email:\ rabin@bose.res.in}
\quad
{\rm and}
\quad
Izumi Tsutsui\footnote{${}^\ddagger$}
{\eightpoint email:\ izumi.tsutsui@kek.jp}
}

\vskip 20pt

{
\baselineskip=13pt
\centerline{\it
Institute of Particle and Nuclear Studies}
\centerline{\it
High Energy Accelerator Research Organization (KEK)}
\centerline{\it Tsukuba 305-0801,  Japan}
}

\vskip 125pt

\abstract{%
{\bf Abstract.}\quad
We generalise the electric-magnetic duality in standard Maxwell theory to
its
non-commutative version. Both space-space and space-time non-commutativity
are
necessary. The duality symmetry is then
extended to a general class of non-commutative gauge theories that goes
beyond
non-commutative electrodynamics. As an application of this symmetry, plane
wave solutions are analysed. Dispersion relations following from these
solutions show 
that general non-commutative gauge theories other than 
electrodynamics
admits two waves with distinct polarisations propagating at different
velocities
in the same direction.
}

}
\ve


\pageheight{23cm}
\pagewidth{15.7cm}
\hcorrection{-1mm}
\magnification= \magstep1
\def\bsk{%
\baselineskip= 14.4pt plus 1pt minus 1pt}
\parskip=5pt plus 1pt minus 1pt
\tolerance 8000
\bsk


\secno=1 \meqno=1

\bigskip
\noindent{\bf 1. Introduction}
\medskip

The old idea [\Snyder] that spatial coordinates need
not commute has recently undergone a revival due to its appearance
in the context of
string theory [\SW]. This has also led to the study of gauge field
theories defined on non-commutative spaces; these are  referred as
non-commutative gauge
theories. Some points of difference, leading to a modification or
generalisation of
the results of conventional gauge theories have been noted.
In particular, in the
non-commutative Maxwell theory, modified dispersion relations
[\GJPP, \Cai], 
violation of the
superposition principle [\BCCMV], non-commuting electric fields
[\Banerjee], {\it etc.}, have been reported.

The objective of this paper is to generalise the
well known electric-magnetic
duality in Maxwell's theory to its
non-commutative counterpart.
The theory is defined
in a non-commutative space-time $x^\mu$ satisfying the algebra
$[ x^\mu, x^\nu] = i\theta^{\mu\nu}$, where
$\theta^{\mu\nu}$ is a real skew 2-index object.
We shall work in the commutative
equivalent of the non-commutative Maxwell
theory provided by the Seiberg-Witten (SW) map [\SW].
Although we perform the analysis up
to the first order in the non-commutative parameter
$\theta$, our results are
expected to hold under more general conditions.
In fact, the duality symmetry
persists (at least to the
first order in $\theta$) for an arbitrary
Lagrangian whose structure is dictated by symmetry arguments and not
restricted by the SW map.

Keeping both space-space $\theta^{ij}$
and space-time $\theta^{0i}$ non-commutativity is
essential for our analysis. Indeed, when we use
the duality symmetry to construct
new solutions, say by taking the plane wave
solutions with nonvanishing
$\theta^{ij}$ but vanishing $\theta^{0i}$ [\GJPP], these are
converted to other plane wave solutions, now with
nonvanishing  $\theta^{0i}$
and vanishing $\theta^{ij}$. This is also consistent
with the known results,
discussed both in the
Lagrangian [\GRS, \RU] and
Hamiltonian formulations [\DY],
that if the original theory is non-commutative Maxwell
theory with only spatial
non-commutativity, the dual theory is a non-commutative
gauge theory with just
space-time non-commutativity.

Our analysis of course goes beyond since
we give the plane wave solutions for
 non-commutativity involving both space and time.
Modified dispersion relations are
obtained, which generalise previous findings [\GJPP] given only for
space-space non-commutativity. 

\secno=2 \meqno=1

\bigskip
\noindent{\bf 2. Duality in the non-commutative Maxwell theory}
\medskip

The non-commutative generalisation of usual Maxwell Lagrange density
involves the
star product of the non-commutative field strength $\widehat F_{\mu\nu}$
obtained from the potential $\widehat A_\mu$
as
$$
\widehat F^{\mu\nu}
= \partial^\mu \widehat A^\nu - \partial^\nu \widehat A^\mu
-ig(\widehat A^\mu \star \widehat A^\nu
- \widehat A_\nu \star \widehat A_\mu),
\eqn\no
$$
where $g = {e\over{\hbar c}}$ and the star product is defined
by
$
(f \star g)(x)= e^{{i\over 2}\theta^{\alpha\beta}\partial_\alpha \partial
'_\beta} f(x) g(x')\big\vert_{x=x'}.
$
The non-commutative version of the free Maxwell Lagrange
density is then given by
$$
{\cal L}
=-\frac{1}{4}\widehat F_{\mu\nu}\star \widehat F^{\mu\nu}.
\eqn\no
$$

This Lagrange
density may be expressed in terms of the field strength
$F_{\mu\nu} = \partial_\mu A_\nu - \partial_\nu A_\mu$
defined from the ordinary  vector potential $A_\mu$ by exploiting
the SW map,\note{%
In this paper we adopt the convention that when the non-commutative
parameter $\theta^{\mu\nu}$ is involved we keep terms only up to first
order in the parameter.}
$$
\eqalign{
\widehat A_{\mu} 
&= A_{\mu} - {1\over 2} \theta^{\alpha\beta}A_{\alpha}
(\partial_\beta A_\mu + F_{\beta\mu}), \cr
\widehat F_{\mu\nu}
&= F_{\mu\nu} + \theta^{\alpha\beta}F_{\alpha\nu} F_{\mu\beta}
- \theta^{\alpha\beta}A_{\alpha}\partial_\beta F_{\mu\nu},
}
\eqn\no
$$
where we have absorbed $g$ into $\theta^{\alpha\beta}$ by scaling.
The outcome is 
$$
{\cal L}
={\cal L}_0
+{\cal L}_1
+{\cal L}_2,
\eqn\lag
$$ 
where ${\cal L}_0
=-\frac{1}{4}F_{\mu\nu}F^{\mu\nu}$ is the ordinary Maxwell Lagrangian and
$$
{\cal L}_1 = 
\frac{1}{8}\theta^{\alpha\beta}F_{\alpha\beta}F_{\mu\nu}F^{\mu\nu},
\qquad
{\cal L}_2 = 
-\frac{1}{2}\theta^{\alpha\beta}F_{\mu\alpha}F_{\nu\beta}F^{\mu\nu},
\eqn\no
$$
are the additional terms manifesting  the effects of non-commutativity. This
was
worked out in [\GRS] and used for a detailed Lagrangian [\GJPP] as well as
Hamiltonian [\Banerjee]
analysis (see also [\BGPSW] for quantization). Using the electric field
$\bE$ and
the magnetic induction field
$\bB$ defined by
$$
E^i=-F^{0i}, \qquad
B^i=-\frac{1}{2}\epsilon^{ijk}F_{jk},
\eqn\ebtdf
$$
and introducing
$$
\varepsilon^i=\theta^{0i}, \qquad
\theta^i=\frac{1}{2}\epsilon^{ijk}\theta_{jk},
\eqn\no
$$
we can rewrite the Lagrangian (\lag) as
$$
{\cal L}
=\frac{1}{2}(\bE^2-\bB^2)
-\frac{1}{2}(\btheta\cdot\bB-\bve\cdot\bE)(\bE^2-\bB^2)
+(\btheta\cdot\bE+\bve\cdot\bB)(\bE\cdot\bB).
\eqn\lageb
$$ 

If we further define the electric displacement field $\bD$ and
the magnetic field $\bH$ by
$$
D^i=\frac{\strut{\textstyle{\partial{\cal L}}}}
             {\strut{\textstyle{\partial(\partial_0 A_i)}}}, \qquad
H^i=\frac{\strut{\textstyle{1}}}{\strut{\textstyle{2}}}\epsilon^{ijk}
        \frac{\strut{\textstyle{\partial{\cal L}}}}
             {\strut{\textstyle{\partial(\partial_j A_k)}}} ,
\eqn\dhdef
$$
which read  
$$
\eqalign{
\bD&=\bE-(\btheta\cdot\bB-\bve\cdot\bE)\bE
+(\btheta\cdot\bE+\bve\cdot\bB)\bB
+(\bE\cdot\bB)\btheta+\frac{1}{2}(\bE^2-\bB^2)\bve, \cr
\bH&=\bB-(\btheta\cdot\bB-\bve\cdot\bE)\bB
-(\btheta\cdot\bE+\bve\cdot\bB)\bE-(\bE\cdot\bB)\bve
+\frac{1}{2}(\bE^2-\bB^2)\btheta,
}
\eqn\dhbe
$$
then the field equations take the Maxwell form:
$$
  \frac{\strut{\textstyle{1}}}{\strut{\textstyle{c}}}
    \frac{\strut{\textstyle{\partial}}}{\strut{\textstyle{\partial t}}}
    \bB+\nabla\times\bE=0, \qquad
   \nabla\cdot\bB=0,
\eqn\bcid
$$
and
$$
   \frac{\strut{\textstyle{1}}}{\strut{\textstyle{c}}}
    \frac{\strut{\textstyle{\partial}}}{\strut{\textstyle{\partial t}}}
    \bD-\nabla\times\bH=0, \qquad
   \nabla\cdot\bD=0.
\eqn\mweq
$$
Note that (\bcid) are just the Bianchi identities while (\mweq) are the
Lagrange equations derived from (\lag).

Because of the presence of both the $\bve$ parameter and
$\btheta$, we find that $\bD$ and $\bH$ in (\dhbe) share
almost the same structure, and this observation leads us to define the
duality
transformation ${\rm D}$ by
$$
   \bE\rightarrow -\bH,\qquad
   \bB\rightarrow \bD,\qquad
   \bve\rightarrow -\btheta,\qquad
   \btheta\rightarrow \bve.
\eqn\dtrs
$$
Obviously, the first two in (\dtrs) render the equations (\bcid)
into (\mweq).  The last two are needed to
ensure the converse, that is, to change (\mweq) back into
(\bcid) so that the entire equations (\bcid)
and (\mweq) are unchanged as a set.  To see how this is done,
we express
$\bE$ and
$\bB$ in terms of $\bD$ and $\bH$ by obtaining the inverse relations for
(\dhbe). Up to ${\cal O}(\theta^{\mu\nu})$,
they are found to be
$$
\eqalign{
\bE&=\bD-(\bve\cdot\bD-\btheta\cdot\bH)\bD
-(\bve\cdot\bH+\btheta\cdot\bD)\bH-(\bH\cdot\bD)\btheta
+\frac{1}{2}(\bH^2-\bD^2)\bve, \cr
\bB&=\bH-(\bve\cdot\bD-\btheta\cdot\bH)\bH
+(\bve\cdot\bH+\btheta\cdot\bD)\bD+(\bH\cdot\bD)\bve
+\frac{1}{2}(\bH^2-\bD^2)\btheta.
}
\eqn\ebhd
$$
It is then easy to see that the
duality transformation (\dtrs) induces
$\bD\rightarrow -\bB$, $\bH\rightarrow \bE$,
and hence (\mweq) is converted to
(\bcid) as claimed.


At this point we point
out that the standard Maxwell electric-magnetic duality $\bE\rightarrow
-\bB$,
$\bB\rightarrow \bE$ is recovered for vanishing non-commutativity
$\theta^{\mu\nu}=0$ in (\dtrs). As is known, this
discrete symmetry gets lifted to a continuous $SO(2)$ symmetry ${\rm
D}(\alpha)$  with $\alpha \in [0, 2\pi)$ by
the following transformation,
$$
\left(\matrix{\bE\cr\bB\cr}\right) \to
\left(\matrix{\bE_{\rm D}\cr\bB_{\rm D}\cr}\right) =
R (\alpha)\left(\matrix{\bE\cr
\bB\cr}\right),
\qquad \hbox{where} \quad
R (\alpha) = 
\left(\matrix{\cos \alpha& -\sin \alpha\cr
\sin \alpha & \cos \alpha\cr}\right).
\eqn\oldmatrx
$$
The standard Maxwell equations,
satisfied by the original variables, are now fulfilled by the rotated dual
variables, and
the discrete electric-magnetic duality is realised by the
choice
$\alpha=\pi/2$, {\it i.e.}, by ${\rm D}({{\pi}\over 2})$.  The $SO(2)$ dual
rotation is actually  a canonical transformation, because the Hamiltonian,
which is just the norm of the vector in the
$\bE-\bB$ space, is preserved by the transformation and, likewise, the
algebra
among the variables ({\it i.e.}, the canonical structure) is preserved.

Likewise, in the non-commutative case $\theta^{\mu\nu} \ne 0$
we have a generalised $SO(2)$
symmetry transformation in which the duality
(\dtrs) is embedded.  Namely, if we introduce the doublets,
$(\bE, \bH)$ and
$(\btheta, \bve)$, and define the dual map ${\rm D}(\alpha)$ by
$$
\left(\matrix{\bE\cr
\bH\cr}\right) \to
\left(\matrix{\bE_{\rm D}\cr
\bH_{\rm D}\cr}\right) = R (\alpha)\left(\matrix{\bE\cr
\bH\cr}\right),
\qquad
\left(\matrix{\bve\cr
\btheta\cr}\right) \to
\left(\matrix{\bve_{\rm D}\cr
\btheta_{\rm D}\cr}\right) = R (\alpha)\left(\matrix{\bve\cr
\btheta\cr}\right),
\eqn\matrx
$$
we find that this induces the same rotation on the doublet $\left(\bD,
\bB\right)^t\to \left(\bD_{\rm D},
\bB_{\rm D}\right)^t = R (\alpha)\left(\bD,
\bB\right)^t$.  We then confirm readily that
the field equations are still satisfied by the dual variables, and
that the discrete dual (\dtrs) is realised by ${\rm D}({{\pi}\over 2})$.
Like in the usual Maxwell case, one also sees that the canonical structure
is
preserved under the rotation.  Combined with the invariance of the
field equations (which implies that the Hamiltonian in the two
sets of variables, original and dual, should agree),
one may regard the rotation as a \lq canonical
transformation\rq, at least heuristically,
even though one cannot explicitly demonstrate this directly, due to the
presence
of higher order time derivatives that arise under
the space-time non-commutativity parameter
$\theta^{0i}$.  For the discrete case ${\rm D}({{\pi}\over 2})$
the duality as a
canonical transformation has been discussed in the
Lagrangian [\GRS, \RU] and
Hamiltonian formulations [\DY].  
The continuous duality rotation symmetry was earlier discussed in 
[\Aschieri], based on a different [\GZ] approach in the context of string
theory.

\secno=3 \meqno=1

\bigskip
\noindent{\bf 3. Duality with arbitrary coefficients
for ${\cal L}_1$ and ${\cal L}_2$}
\medskip

The duality symmetry can be observed even with
arbitrary coefficients for ${\cal L}_1$ and ${\cal L}_2$ in (\lag).
Namely, we consider a system governed by the Lagrangian,\note{%
Since $\theta$ can be scaled freely, one can always normalize the
coefficient for ${\cal L}_1$ (or ${\cal L}_2$) even if one starts with
arbitrary
coefficients for both ${\cal L}_1$ and ${\cal L}_2$.
} 
$$
{\cal L}
={\cal L}_0
+{\cal L}_1
+a{\cal L}_2,
\eqn\laggen
$$ 
with an arbitrary real parameter $a$.
Note that this is the most general
Lagrangian, up to ${\cal O}(\theta^{\mu\nu})$,
constructed out of
$F^{\mu\nu}$ and
$\theta^{\mu\nu}$,
that has similar Lorentz
tranformation properties as the
non-commutative Maxwell theory (\lag) which
is recovered by choosing $a = 1$.
The field equations (\bcid) and
(\mweq) still hold for any $a$
with $\bD$
and $\bH$  defined by (\dhdef) which are now $a$-dependent,
$$
\eqalign{
\bD
&=\bigl[1-(2a-1)(\btheta\cdot\bB
-\bve\cdot\bE)\bigr] {\bE}
 +a(\btheta\cdot \bE
+\bve\cdot\bB)\bB\cr
&\qquad\qquad
+a(\bE\cdot\bB)\btheta
  +\frac{1}{2}(2a-1)(\bE^2-\bB^2)\bve,\cr
  \bH&=\bigl[1-(2a-1)(\btheta\cdot\bB
-\bve\cdot\bE)\bigr]\bB
-a(\btheta\cdot\bE
+\bve\cdot\bB)\bE\cr
&\qquad\qquad
-a(\bE\cdot\bB)\bve
  +\frac{1}{2}(2a-1)\bigl(\bE^2-\bB^2\bigr)\btheta.
}
\eqn\dheb
$$
The inverse relations of these are then given by
$$
\eqalign{
\bE&=\bigl[1+(2a-1)(\btheta\cdot\bH
-\bve\cdot\bD)\bigr]\bD
-a(\btheta\cdot\bD
+\bve\cdot\bH)\bH
\cr
&\qquad\qquad
-a(\bD\cdot\bH)\btheta
  -\frac{1}{2}(2a-1)(\bD^2-\bH^2)\bve, \cr
\bB &=\bigl[1+(2a-1)(\btheta\cdot\bH
-\bve\cdot\bD)\bigr]\bH
+a(\btheta\cdot\bD
+\bve\cdot\bH)\bD
\cr
&\qquad\qquad
+a(\bD\cdot\bH)\bve
  -\frac{1}{2}(2a-1)\bigl(\bD^2-\bH^2\bigr)\btheta.
}
\eqn\exebdh
$$
As in the non-commutative case, one can observe that
the duality transformation (\dtrs) --- which now depends on $a$ --- induces
$\bD\rightarrow -\bB$, $\bH\rightarrow \bE$, and hence it  preserves the
field
equations (\bcid) and (\mweq) as a set.  One therefore sees that, as long as
the
duality is concerned, the non-commutative Maxwell theory occupies no special
position
among the other theories defined by (\laggen).

Our argument on the duality can be made concise if we use
$F^{\mu\nu}$ and $\theta^{\mu\nu}$ and introduce
$$
G^{\mu\nu}=-\frac{\partial{\cal L}}{\partial(\partial_\mu A_\nu)},
\eqn\no
$$
which is the tensor for $\bD$ and $\bH$, {\it i.e.},
$$
D^i=-G^{0i}, \qquad H^i=-\frac{1}{2}\epsilon^{ijk}G_{jk}.
\eqn\no
$$
In terms of $F^{\mu\nu}$, the tensor $G^{\mu\nu}$ is
written as
$$
\eqalign{
G^{\mu\nu}
&=F^{\mu\nu}-\frac{1}{4}(\theta^{\mu\nu}F_{\alpha\beta}F^{\alpha\beta}
+2\theta^{\alpha\beta}F_{\alpha\beta}F^{\mu\nu})
\cr
&\qquad
+a\Big[(\theta^{\mu\alpha}F^{\nu\beta}
-\theta^{\nu\alpha}F^{\mu\beta})F_{\alpha\beta}
+\theta_{\alpha\beta}F^{\mu\alpha}F^{\nu\beta}\Big],
}
\eqn\no
$$
which admits the inverse,
$$
\eqalign{
F^{\mu\nu}
&=G^{\mu\nu}+\frac{1}{4}(\theta^{\mu\nu}G_{\alpha\beta}G^{\alpha\beta}
+2\theta^{\alpha\beta}G_{\alpha\beta}G^{\mu\nu})
\cr
&\qquad -a\Big[(\theta^{\mu\alpha}G^{\nu\beta}
-\theta^{\nu\alpha}G^{\mu\beta})G_{\alpha\beta}
+\theta_{\alpha\beta}G^{\mu\alpha}G^{\nu\beta}\Big].
}
\eqn\no
$$
In this tensorial notation, the field equations (\bcid) and (\mweq)
take the simple form,
$$
\partial_{\mu}{}^\ast F^{\mu\nu}=0,
\qquad
 \partial_{\mu}G^{\mu\nu}=0.
\eqn\emt
$$
where we have used the dual of the field strength,
${}^\ast F^{\mu\nu}
=\frac{1}{2}\epsilon^{\mu\nu\alpha\beta}F_{\alpha\beta}$.
In terms of the dual tensors ${}^\ast G^{\mu\nu}$ and
${}^\ast\theta^{\mu\nu}$ similarly defined,
the duality transformation (\dtrs) takes the simple form,
$$
 F^{\mu\nu} \rightarrow -{}^\ast G^{\mu\nu}, \qquad
  \theta^{\mu\nu} \rightarrow -{}^\ast\theta^{\mu\nu}.
\eqn\dtrst
$$
The duality transformation (\dtrst)
induces 
$G^{\mu\nu} \rightarrow -{}^\ast F^{\mu\nu}$,
confirming that the field equations (\emt) are indeed preserved
under  (\dtrst), thereby providing another example of  
electric-magnetic duality.

\secno=4 \meqno=1

\bigskip
\noindent{\bf 4. Plane wave solutions}
\medskip

We now utilize the discrete duality transformation ${\rm D}({{\pi}\over 2})$
to
generate new solutions from known
plane wave solutions (more general solutions for ${\rm
D}(\alpha)$ may be obtained by the rotation once the discrete dual solution
is obtained).  To this end, we first recall that for
$$
 \bve=0, \qquad \btheta \ne 0,
\eqn\no
$$
the non-commutative Maxwell theory (\lag) admits the plane wave
solution [\GJPP],
$$
\bE=\bE(\omega t-\bk\cdot\br), \qquad
\bB=\bkp\times\bE+\bb,
\eqn\pws
$$
where $\bkp=c\bk/\omega$ and $\bb$ represents a constant background.  The
constants
$\omega$ and
$k =
\vert
\bk\vert$ fulfill the dispersion relation,
$$
 \omega=ck(1-\btheta_T\cdot\bb_T),
\eqn\no
$$
with $\btheta_T$ (and similarly $\bb_T$) being the transverse part
of 
$\btheta$ (and $\bb$)
with respect to the wave vector $\bk$.  No condition arises for the
transverse part $\bE_T$ of the electric field $\bE = \bE_T + E_L \hat{\bkp}$
(we put a hat on normalized vectors as $\hat{\bkp} = {\bkp}/\vert
\bkp\vert$) but the longitudinal part
is subject to the condition,
$$
 E_L=-(\bb_T\cdot\bE_T)\theta_L-(\btheta_T\cdot\bE_T)b_L.
\eqn\plsl
$$
Thus, the plane wave possesses two degrees of freedom for polarisation as
in the standard Maxwell theory but the polarisation is no longer
transverse in general.   This solution may be characterized by the
property that no background field contributes to $\bE$ and
$\bD$ while it does to $\bH$ as well as $\bB$.  In fact, from (\dhbe)
combined
with  (\pws) and (\plsl) one finds
that $\bH$ possesses the constant background,
$$
 \bh=\big[1-(\btheta\cdot\bb)\big]\bb-\frac{1}{2}\vert \bb\vert^2\btheta.
\eqn\hsol
$$ 

One can obtain a new plane wave solution from this solution by performing
the 
discrete duality transformation (\dtrs).  Since the duality involves the
interchange of the parameters $\btheta$ and $\bve$, the new solution is
valid
for 
$$
 \bve \ne 0, \qquad \btheta = 0,
\eqn\no
$$
and takes the form,
$$
 \bH=\bH(\omega t-\bk\cdot\br), \qquad
\bD= - \bkp\times\bH+\bd,
\eqn\pwsd
$$
with a constant background $\bd$.  The dispersion
relation now reads
$$
 \omega=ck(1-\bve_T\cdot\bd_T).
\eqn\ddr
$$
Analogously to the electric field $\bE$ in the previous case, the
magnetic field $\bH$ admits two
polarisation degrees of freedom in
the transverse part $\bH_T$ while the
longitudinal part is determined as
$$
 H_L=-(\bd_T\cdot\bH_T)\varepsilon_L-(\bve_T\cdot\bH_T)d_L.
\eqn\no
$$
In this solution, $\bE$ has a
constant background $\be$ given by
$$
\be =\big[1-(\bve\cdot\bd)\big]\bd-\frac{1}{2}\vert \bd\vert^2\bve.
\eqn\no
$$

The above argument can be carried over to
the general case where both
$\btheta$ and
$\bve$ are nonvanishing and also for theories with an
arbitrary coefficient $a$.
As before, if we start with the plane wave ansatz (\pws)
for $\bE$ and
$\bB$, then from the field equations (\bcid) and (\mweq)
we find that the
longitudinal part of the electric field is determined as
$$
\eqalign{
E_L
&=-a(\bb_T\cdot\bE_T)\theta_L-a\big(\btheta_T\cdot\bE_T
+\bve_T\cdot(\bkp\times\bE_T)\big)b_L \cr
&\qquad
+(2a-1)\big(\bb_T\cdot(\bkp\times\bE_T)\big)\varepsilon_L.
}
\eqn\elnp
$$
On the other hand, the transverse part
$\bE_T$ must fulfill the condition,
$$
\left(M^{ij} - \lambda\, \delta^{ij}\right) E_T^j = 0,
\eqn\etcond
$$
with
$$
\eqalign{
M^{ij}
&=(a-1)\Big[\theta_T^ib_T^j+\theta_T^jb_T^i
+(\epsilon^{ik}\varepsilon_T^j
+\epsilon^{jk}\varepsilon_T^i)
b_T^k\Big], \cr
\lambda
&=1-\kappa^2+2(2a-1)(\btheta_T\cdot\bb_T)
+2a(\bve_T\cdot(\hat{\bkp}\times\bb_T)),
}
\eqn\no
$$
where we have chosen the third spacial ($z-$)axis in the direction of
$\hat\bkp$ so that $i, j$ run over 1, 2 and used
the antisymmetric tensor
$\epsilon^{ij}$ with
$\epsilon^{12} = 1$.  From (\etcond)
it follows that, unless the matrix $(M^{ij} - \lambda\,
\delta^{ij})$ vanishes identically, the polarisation in the transverse part
$\bE_T$ is determined by the non-commutative parameters $\btheta$, $\bve$
and the coefficient $a$ together with $\bkp$.  Explicitly,
we find that $\bE_T$ points to the specific directions,
$$
\eqalign{
\bE_T
&\propto\bigg[\frac{\hat{\bve}_T\cdot\hat{\btheta}_T}{\hat{\bb}_T\cdot
(\hat{\bkp}\times\hat{\btheta}_T)}-\frac{\hat{\bve}_T\cdot
(\hat{\bkp}\times\hat{\btheta}_T)-\alpha\pm\sqrt{A}}
{1-\hat{\bb}_T\cdot\hat{\btheta}_T}\bigg]
\hat{\bb}_T \cr
&\qquad+\bigg[\frac{\hat{\bve}_T\cdot\hat{\btheta}_T}{\hat{\bb}_T
\cdot(\hat{\bkp}\times\hat{\btheta}_T)}
+\frac{\hat{\bve}_T\cdot(\hat{\bkp}\times\hat{\btheta}_T)-\alpha\pm\sqrt{A}}{1-\hat{\bb}_T
\cdot\hat{\btheta}_T}\bigg]\hat{\btheta}_T,
}
\eqn\etrp
$$
with
$$
\eqalign{
\alpha
&= \theta_T/\varepsilon_T, \cr
A
&= 1+\alpha^2+2\alpha(\hat{\bve}_T\cdot(\hat{\bkp}\times\hat{\btheta}_T)).
}
\eqn\no
$$
Once $\bE$ is known, $\bB$ is determined uniquely from (\pws). Using
(\dheb), one can also obtain $\bH$, which has the oscillatory
longitudinal part,
$$
\eqalign{
H_L
&=-a(\bb_T\cdot\bE_T)\varepsilon_L
-(2a-1)
\big(\btheta_T\cdot(\bkp\times\bE_T)-\bve_T\cdot\bE_T \big)b_L \cr
&\qquad
-(2a-1)\big(\bb_T\cdot(\bkp\times\bE_T)\big)\theta_L.
}
\eqn\hlnp
$$
The constant background of $\bH$ is given by
$$
\bh=\big[1-(2a-1)(\btheta\cdot\bb)\big]\bb-\frac{1}{2}(2a-1)
\vert \bb\vert^2 \btheta,
\eqn\hcst
$$
which is nonvanishing in the limit $\theta^{\mu\nu} \to 0$,
whereas the constant background of $\bD$ is
$$
\bd= a(\bve\cdot\bb)\bb-\frac{1}{2}(2a-1)
\vert \bb\vert^2 \bve,
\eqn\dcst
$$
which vanishes in the limit.

A nonvanishing
$\bE_T$ requires
$\det (M^{ij} - \lambda\, \delta^{ij}) = 0$,
and from this one obtains the dispersion relation
$$
\omega=ck\Big[1-\frac{1}{2}(3a-1)\big(\btheta_T\cdot\bb_T
+\bve_T\cdot(\hat{\bkp}\times\bb_T)\big)\pm\frac{a-1}{2}b_T\varepsilon_T\sqrt{A}\Big]
\eqn\dprl
$$
corresponding to the two expressions for $\bE_T$ given in (\etrp).
Observe that, unless $A = 0$,
there are two solutions for (\dprl), which implies that the two waves with
distinct polarisations  propagate at different velocities even in the same
direction.
Note that, as mentioned earlier,  the two
polarisations can be in arbitrary directions if $M^{ij} = \lambda\,
\delta^{ij}$. Without attempting a general solution for this equation,
let us remark
that this condition is trivially satisfied for non-commutative
electrodynamics $(a=1)$. 
In this case both $M^{ij}$ and $\lambda$
vanish identically and, if further $\bve = 0$, the
solution reduces to the one discussed in [\GJPP].

If we implement the discrete duality (\dtrs) to
the above solution, we obtain a new solution with the ansatz
(\pwsd).
The longitudinal and transverse parts of $\bH$ are gained directly from
(\elnp) and (\etrp) by the duality map.
The transverse parts of $\bH$ is
$$
\eqalign{
\bH_T
&\propto \Bigg[\frac{\hat{\bve}_T\cdot\hat{\btheta}_T}
{\hat{\bd}_T\cdot(\hat{\bkp}\times\hat{\bve}_T)}
+\frac{\hat{\bve}_T\cdot
(\hat{\bkp}\times\hat{\btheta}_T)-\tilde{\alpha}\pm\sqrt{\tilde{A}}}
{1-\hat{\bd}_T\cdot\hat{\bve}_T}\Bigg]\hat{\bd}_T \cr
&\qquad+\Bigg[\frac{\hat{\bve}_T\cdot\hat{\btheta}_T}
{\hat{\bd}_T\cdot(\hat{\bkp}\times\hat{\bve}_T)}
-\frac{\hat{\bve}_T\cdot
(\hat{\bkp}\times\hat{\btheta}_T)-\tilde{\alpha}\pm\sqrt{\tilde{A}}}
{1-\hat{\bd}_T\cdot\hat{\bve}_T}\Bigg]\hat{\bve}_T,
}
\eqn\tbcdn
$$
where 
$$
\eqalign{
\tilde{\alpha}
&= \varepsilon_T/\theta_T, \cr
\tilde{A}
&= 1+\tilde{\alpha}^2+2\tilde{\alpha}(\hat{\varepsilon}_T
\times\hat{\theta}_T)).
}
\eqn\no
$$
The longitudinal part of $\bE$ can be found by
applying the duality map to (\hlnp) as
$$ 
\eqalign{
E_L
&=-a(\bd_T\cdot\bH_T)\theta_L+(2a-1)
\big(\bve_T\cdot(\hat{\bkp}\times\bH_T)+\btheta_T\cdot\bH_T\big)d_L \cr
&\qquad +(2a-1)\big(\bd_T\cdot(\hat{\bkp}\times\bH_T)\big)\varepsilon_L.
}
\eqn\no
$$

The duality map (\dtrs) is also used to obtain
the dispersion relation,
$$
\omega=ck\Big[1-\frac{3a-1}{2}
\big(\bve_T\cdot\bd_T-\btheta_T\cdot(\hat{\bkp}
\times\bd_T)\big)\pm\frac{a-1}{2}d_T\theta_T\sqrt{\tilde{A}}\Big],
\eqn\no
$$
directly from (\dprl).
Analogously to (\etcond), $\bH_T$ is subject to
$(\tilde M^{ij} - \tilde\lambda\, \delta^{ij})H_T^j=0$ (which could also be
used
to obtain (\tbcdn)) with
$$
\eqalign{
\tilde M^{ij}
&=(a-1)\big[d_T^i\varepsilon_T^j+d_T^j\varepsilon_T^i
-(\epsilon^{ik}\theta_T^j+\epsilon^{jk}\theta_T^i)d_T^k\big], \cr
\tilde \lambda
&=1-\kappa^2+2(2a-1)(\bve_T\cdot\bd_T)-2a\big(\btheta_T
\cdot(\hat{\bkp}\times\bd_T)\big).
}
\eqn\mtcd
$$

We have seen, therefore, that the non-commutative electrodynamics is
distinguished in that it admits a plane wave solution with arbitrary
polarisation
in the transverse part (from which the longitudinal part is determined) 
under any choice of the non-commutative parameter $\theta^{\mu\nu}$.  
This should be contrasted to the case
$a \ne 1$ where there is  a severe restriction on
the polarisation as a consequence of
the non-commutativity for plane wave solutions with a generic constant
background.  
This result is valid no matter how small the constant background and/or the
non-commutativity parameters are, and this may be used to provide some
stringent 
constraints on the value of
$\bve$ or $\btheta$ in laboratory tests.
The singularity of the plane wave solution in the limit $\theta^{\mu\nu} \to
0$
suggests, however, that there may be other
\lq quasi plane wave\rq{} solutions which do not assume the plane wave
ansatz for
$\theta^{\mu\nu} \ne 0$ but nonetheless reduce to the usual plane wave in
the limit
$\theta^{\mu\nu} \to 0$.

\secno=5 \meqno=1

\bigskip
\noindent{\bf 5. Conclusion}
\medskip

We have generalised the notion of electric-magnetic duality to
non-commutative
gauge theories, expanded as a power series in terms of the non-commutativity
parameter. The coefficients of the terms additional to the usual Maxwell
piece
were completely arbitrary and not constrained by the Seiberg-Witten map.
Naturally, non-commutative electrodynamics obtained by an application of
this
map manifested this duality, since it just meant the fixing of the arbitrary
parameters. The explicit map between the electric and magnetic variables was
provided. Our results were compatible with previous analysis on $S$-duality
in
non-commutative theories [\GRS, \RU, \DY,
\Lozano]. 
As an
application, we worked out the plane wave solutions under various
conditions.
Dispersion relations obtained from these solutions led to certain intriguing
possibilities; in particular, the propagation of two waves, with distinct
polarisations, at different velocities in the same direction.

\bigskip
\noindent{\bf Acknowledgements}
\medskip

One of the authors (RB) thanks the JSPS for providing the support which made
this
collaboration possible. He also thanks the members of the Theory group at
KEK, for
their hospitality.
This work has been supported in part by
the Grant-in-Aid for Scientific
Research on Priority Areas (No.~13135206) by
the Japanese
Ministry of
Education, Science, Sports and Culture.

\bigskip
\noindent{\bf Note added}
\medskip

In the previous version we had an error  in
deriving our dispersion relations leading to minor modifications in eqs. (4.13),  (4.14),  (4.15) and (4.19).   This however does not lead to any change in our conclusions except
for the fact that noncommutative electrodynamics ({\it i.e.}, the case $a=1$) 
admits arbitray polarisations even for $\bve \ne 0$.  All necessary corrections are implemented in this version, and now our observation on polarisations is in agreement with the conclusion of the paper,   J. Zahn, {\it Phys. Rev.} {\bf D70} (2004) 107704.    We thank the author for communicating with us about the error.

\baselineskip= 15.5pt plus 1pt minus 1pt
\parskip=5pt plus 1pt minus 1pt
\tolerance 8000
\vfill\eject\immediate\closeout\reffile
\centerline{{\bf References}}\bigskip\frenchspacing%
\input refs.tmp\vfill\eject\nonfrenchspacing

\bye